\RequirePackage{fix-cm}
\documentclass[twocolumn,epjc3]{svjour3}  
\RequirePackage{graphicx}
\usepackage{tikz}
\RequirePackage{mathptmx}
\RequirePackage{flushend}
\usepackage[english]{babel}
\usepackage{amsmath}
\usepackage{mathtools}
\usepackage{siunitx}
\usepackage{comment}
\usepackage{tabularx}
\usepackage[version=4]{mhchem}
\usepackage[font={small},labelfont=bf,labelsep=quad]{caption}
\usepackage[font={normal}]{subcaption}
\usepackage{cite}
\usepackage{arydshln}
\usepackage[switch]{lineno}
\usepackage{bm}
\RequirePackage[colorlinks,citecolor=blue,urlcolor=blue,linkcolor=blue]{hyperref}
\RequirePackage[numbers,sort&compress]{natbib}
\usepackage{orcidlink}

\DeclareMathAlphabet{\mathcal}{OMS}{cmsy}{m}{n} 

\newcommand{\zurich}{Physik-Institut, University of Z\"urich,  Winterthurerstrasse 190, 8057  Z\"urich, Switzerland}
\newcommand{\frascati}{INFN, Laboratori Nazionali di Frascati, Via E. Fermi 54, I-00044 Roma, Italy }
\newcommand{\cref}{Centro Ricerche Enrico Fermi---Museo Storico della Fisica e Centro Studi e Ricerche ``Enrico Fermi'', 00184 Rome, Italy}
\newcommand{\heidelberg}{Max-Planck-Institut f\"ur Kernphysik, 69117 Heidelberg, Germany}

\journalname{Eur. Phys. J. C}

\begin{document}
\sloppy

\title{Search for Pauli Exclusion Principle Violations with Gator at LNGS}

\author{
        L.~Baudis\thanksref{UZH}\,\orcidlink{0000-0003-4710-1768}\and
        R.~Biondi\thanksref{MPIK}\,\orcidlink{0000-0002-6622-8740}\and
        A.~Bismark\thanksref{UZH, e1}\,\orcidlink{0000-0002-0574-4303}\and
        A.~Clozza\thanksref{INFN}\,\orcidlink{0000-0003-2133-1725}\and
        C.~Curceanu\thanksref{INFN}\,\orcidlink{0000-0002-1990-0127}\and
        M.~Galloway\thanksref{UZH}\,\orcidlink{0000-0002-8323-9564}\and
        F.~Napolitano\thanksref{INFN, e2}\,\orcidlink{0000-0002-8686-5923}\and
        F.~Piastra\thanksref{UZH}\,\orcidlink{0000-0001-8848-5089}\and
        K.~Piscicchia\thanksref{CREF,INFN}\,\orcidlink{0000-0001-6879-452X}\and
        A.~Porcelli\thanksref{CREF,INFN}\,\orcidlink{0000-0002-3220-6295}\and
        D.~Ram\'irez~Garc\'ia\thanksref{UZH}\,\orcidlink{0000-0002-5896-2697}
 } 
 
\thankstext{e1}{e-mail: alexander.bismark@physik.uzh.ch}
\thankstext{e2}{e-mail: napolitano.fabrizio@lnf.infn.it}

\institute{
\zurich\label{UZH}
\and
\heidelberg\label{MPIK}
\and
\frascati\label{INFN}
\and
\cref\label{CREF}
} 

\date{Received: date / Accepted: date}

\maketitle

\begin{abstract}

The Pauli Exclusion Principle (PEP) appears from fundamental symmetries in quantum field theories, but its physical origin is still to be understood. High-precision experimental searches for small PEP violations permit testing key assumptions of the Standard Model with high sensitivity. We report on a dedicated measurement with Gator, a low-background, high-purity germanium detector operated at the Laboratori Nazionali del Gran Sasso, aimed at testing PEP-violating atomic transitions in lead. The experimental technique, relying on forming a new symmetry state by introducing electrons into the pre-existing electron system through a direct current, satisfies the conditions of the Messiah-Greenberg superselection rule. No PEP violation has been observed, and an upper limit on the PEP violation probability of $\beta^2/2 < \SI{4.8e-29}{}$ (\SI{90}{\percent} CL) is set. This improves the previous constraint from a comparable measurement by more than one order of magnitude.
\end{abstract}

\section{Introduction}
\label{sec:intoduction}

The Pauli Exclusion Principle is so deeply entrenched in our understanding of quantum theory that \mbox{high-sensitivity} experimental tests were not performed until the end of the last century to constrain the limits of its validity \cite{goldhaber1948identification,ramberg1990experimental,deilamian1995search,de1996test,hilborn1996spectroscopic,modugno1998search}. 
The strength of the principle is supported by several features which are direct consequences of the PEP. Dyson, Lenard, Lieb, and Thirring investigated, e.g., the stability of an assembly of $N$ identical particles \cite{Dyson_1967,Lenard_1968,lieb1976stability,lieb2001bound}. They found that there exists no minimum
for the binding energy of a relativistic system of identical bosons, whose fate is to collapse, in contrast to systems obeying the Fermi-Dirac statistics like neutron stars, which are stable due to the degeneracy
pressure exerted by the identical fermions obeying the Pauli principle. Fermi discussed the possibility of ``slightly non-identical'' electrons,   \cite{fermi,edoardo}, concluding that the consequent modifications of the atomic properties would be evident after billions of years of their existence. Moreover, if two different types of electrons existed, modification of pair production processes would have been observed, e.g., in Bhabha scattering produced at electron-positron colliders \cite{amado1980comments}.  

Nevertheless, a renewed interest emerged in the last decades for the development of theories embedding violation of statistics and for their experimental investigation. This is motivated by the profound link of the spin-statistics connection to fundamental assumptions of the Standard Model (SM) of particle physics \cite{greenberg2000theories,mavromatos2017models}, such as Lorentz invariance and CPT symmetry. Hence, experimental tests of spin-statistics can provide extreme sensitivity tests of the SM pillars. 

Several attempts to build consistent local quantum field theories, which permit interpolation between Fermi and Bose statistics \cite{gentile,green,dellantonio,ik1988,IGNATIEV20062090,okun,govorkov1989can}, culminated with the development of the ``quon model'' \cite{Greenberg:1987aa,greenberg1991particles}. The quon algebra is determined by the relation 
\begin{equation}
    a_k a_l^{\dagger} - q a_l^{\dagger} a_k = \delta_{k,l},
\end{equation}
with $a$ and $a^{\dagger}$ representing the creation and annihilation operators. For $q$ ranging between -1 and 1, all representations of the symmetric group occur; $q = -1$ corresponds to the totally antisymmetric representation (Fermi-Dirac statistics), while for $q = 1$, the totally symmetric representation is recovered (Bose-Einstein statistics). If a small violation of Fermi statistics can occur with probability $\beta^2/2$, then the concrete interpretation of $q$ is: 
\begin{equation}
    \beta^2 = 1+q
\end{equation}
i.e., $\beta^2$ is the coefficient of the anomalous component of the two-identical-fermions density matrix,
\begin{equation}
    \rho_2 = (1-\beta^2) \, \rho_a + \beta^2 \, \rho_s,
\end{equation}
where $\rho_{a(s)}$ is the antisymmetric (symmetric) form.

\begin{figure}[htb]
	\centering
	\includegraphics[width=0.99\linewidth]{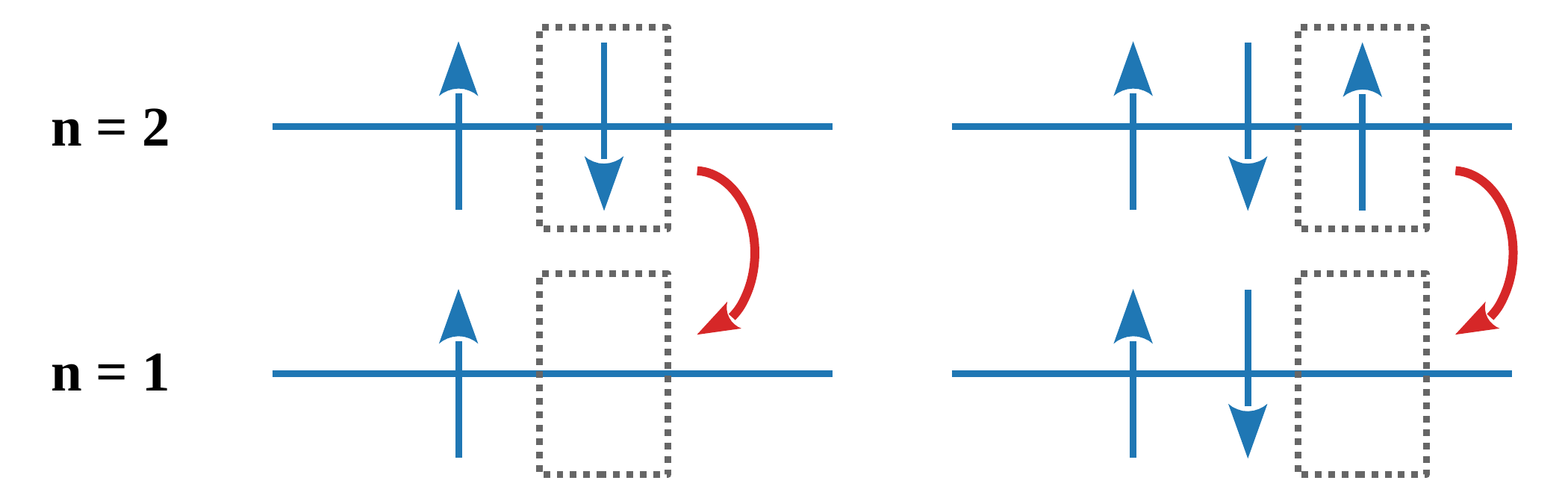}
	\caption{Schematic illustration of a standard (left) and non-Paulian (right) atomic transition from the $L$-shell to the $K$-shell.} 
	\label{fig:pep_schematic}
\end{figure}

Even though a small probability would permit mixed symmetry components in wave functions that are otherwise antisymmetric, the Hamiltonian cannot change the symmetry state of any multi-particle wave function, since it must
be totally symmetric in the dynamical variables of the identical particles. The consequence is known as the Messiah-Greenberg (MG) superselection rule \cite{superselection,amado1980comments} and stipulates to test $\beta^2/2$ by looking for transitions among anomalous symmetry states. This is realized by introducing new fermions in a pre-existing system of identical fermions and probing the newly formed symmetry state. An example, at the basis of the experiment presented in this work, is given by an atomic system for which the newly formed symmetry state of the electrons is mixed. In this case, the $K$-shell may host three electrons, and electrons from higher shells would perform transitions to the fundamental level by respecting the standard branching ratios, as illustrated in \autoref{fig:pep_schematic}.

In Ref.~\cite{elliott2012improved} the definition of 
``new fermion'' is discussed, and experimental tests of PEP are classified into three categories: 

\begin{enumerate}
    \item Type I exploits target fermions that have not previously interacted with any other fermion. A prototype type I experiment was carried out in Ref.~\cite{goldhaber1948identification}, looking for PEP-violating captures of ${}^{14}$C $\beta$-rays by Pb atoms. The corresponding limits on  $\beta^2/2$ were recently improved in Refs.~\cite{elliott2012improved,Majorana:2022mrm}.
    \item Type II measurements exploit target fermions that never previously
interacted with a given system of identical fermions.
The experiment described in this work belongs to this class, which was pioneered by Ramberg and Snow \cite{ramberg1990experimental} following a suggestion of Greenberg and Mohapatra \cite{Greenberg:1987aa}. The method consists of injecting new electrons by means of a direct current in a conductive target and looking for a difference in the X-ray emission with current on and off. K$_{\alpha_{1,2}}$ PEP violating transitions are searched for, whose characteristic signature is a shift downwards in energy, with respect to the standard transitions, due to the additional screening of the nucleus provided by the second electron in the 1$s$ level. In Ref.~\cite{ramberg1990experimental}, the PEP violation in copper (Cu) atoms is studied, and the strongest limit on $\beta^2/2$ for Cu was recently obtained in Ref.~\cite{napolitano2022testing}. Given the importance of testing the PEP violation probability for various elements (see Ref.~\cite{okun}), the experiment was performed in Ref.~\cite{elliott2012improved} by using a Pb target and the limit $\beta^2/2< \SI{1.5e-27}{}$ was obtained.
    \item In type III experiments, no new fermions are introduced. Instead, the target fermions already belong to the fermions system under study (see, e.g., \cite{bernabei2009new}). All type III experiments violate the MG superselection rule and cannot be used to test the quon theory, but they were recently shown to set stringent bounds on the spin-statistics deformation induced by Non-Commutative Quantum Gravity models (see, e.g., \cite{addazi2018testing,piscicchia2022strongest,piscicchia2023first}).    
\end{enumerate}
 
In this work, we report the results obtained in a dedicated measurement performed with the Gator detector. The aim is to test $\beta^2/2$ for Pb atoms in a type II experiment with high sensitivity.

The manuscript is organized as follows: Section~\ref{sec:experimental_setup} describes the experimental setup, Sections~\ref{sec:measurements} and \ref{sec:results_discussion} discuss data taking and analysis, and Section \ref{sec:conclusions} gives concluding remarks and an outlook. 

\section{The experimental setup}
\label{sec:experimental_setup}

The Pauli Exclusion Principle violation studies presented in this work were conducted at the Gator high-purity germanium (HPGe) detector facility. This facility is operated underground at the Laboratori Nazionali del Gran Sasso (LNGS) of INFN in Italy at an average depth of 3600 meter water equivalent~\cite{Baudis:2011am,Araujo:2022kip}. It is primarily used for high-sensitivity $\gamma$-ray spectrometry for material radioassay in ultra-low background, rare-event search experiments, such as XENONnT~\cite{XENON:2021mrg}, LEGEND-200, as well as  DARWIN/XLZD~\cite{DARWIN:2016hyl,Baudis:2024jnk} and LEGEND-1000~\cite{LEGEND:2021bnm}. Owing to its low background rate, Gator also offers sensitivity to X-rays from atomic transitions in Pb investigated in this work. 

\subsection{The Gator HPGe detector facility}

\begin{figure}[htb]
	\centering
	\includegraphics[trim={100 50 100 100},clip,width=0.99\linewidth]{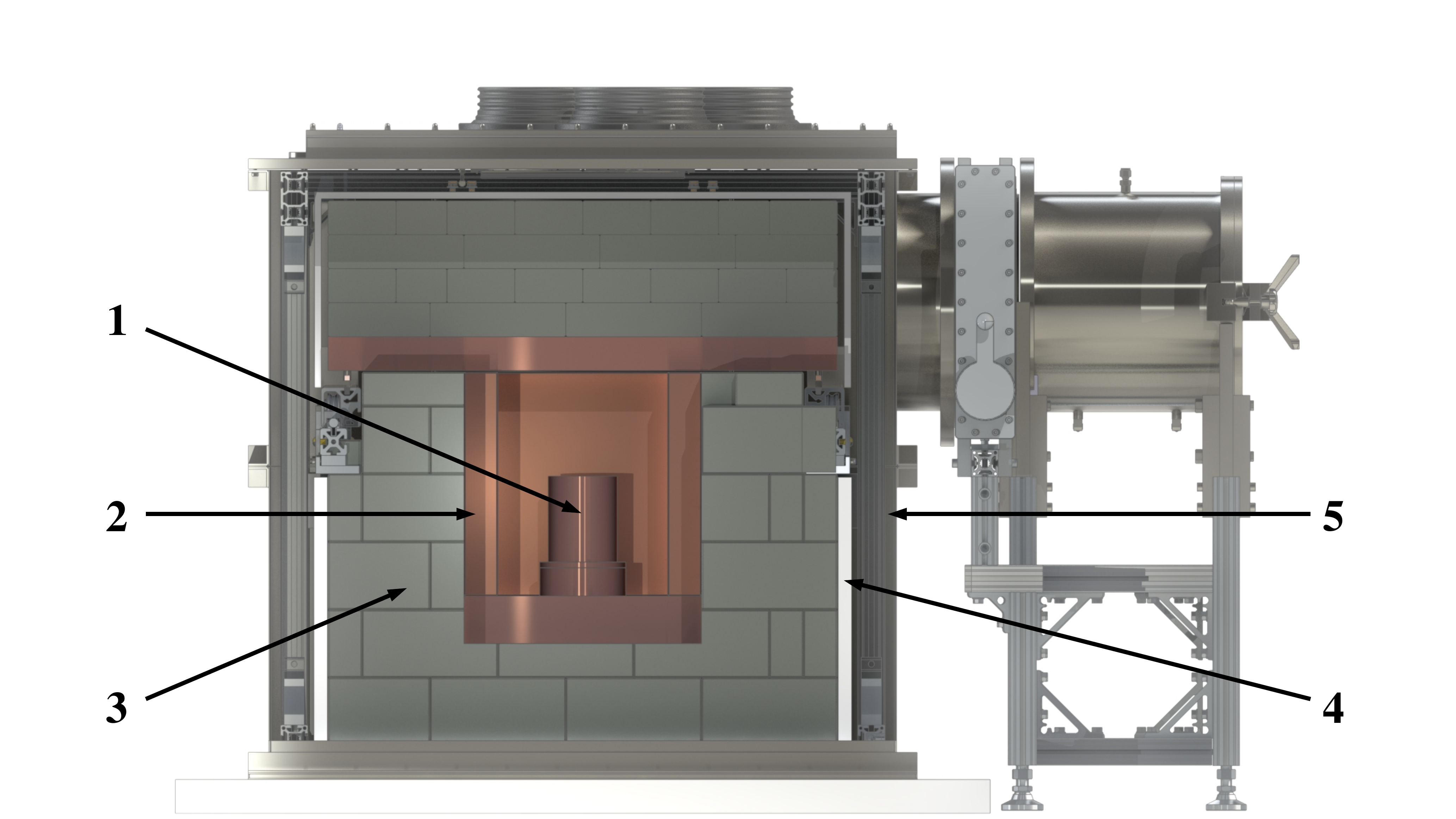}
	\caption{Schematic view of the Gator setup. A Cu cryostat~(1) houses the HPGe crystal within the sample cavity, which is formed by surrounding layers of low-background Cu~(2), Pb~(3), and polyethylene~(4). The stainless steel enclosure~(5) is continuously purged with nitrogen gas for radon suppression. Figure adapted from~\cite{Araujo:2022kip}.} 
	\label{fig:gator_facility}
\end{figure}

At its core, Gator deploys a \SI{2.2}{kg}, p-type coaxial HPGe detector with a relative efficiency of  \SI{100.5}{\percent}~\cite{Baudis:2011am}. The HPGe crystal is housed in an ultra-low background, oxygen-free Cu cryostat and surrounded by a passive shield made of layers of Cu, Pb, and polyethylene. The sample cavity has inner dimensions of $25 \times 25 \times \SI{33}{cm^3}$ and is continuously purged with gaseous nitrogen for radon suppression. The detector is operated at $\sim \SI{90}{K}$, with cooling provided by a Cu coldfinger immersed in liquid nitrogen. Glove ports in an acrylic plate above the all-engrossing stainless steel enclosure and a lateral load-lock chamber facilitate the sample loading for material radioassay measurements. The main components of the facility are illustrated in \autoref{fig:gator_facility}. The integrated background rate inside the empty sample cavity is \SI{82.0(7)}{counts/(kg.day)} in the energy range \SIrange{100}{2700}{keV} and \SI{4.4(3)}{counts/(kg.day)} in the energy region \SIrange{65}{90}{keV} around the investigated Pb X-rays. A detailed description of the facility previous to changes made for the PEP violation studies can be found in Refs.~\cite{Baudis:2011am,Araujo:2022kip}.

\subsection{The PEP violations study setup}

To implement the Ramberg-Snow technique for the Pauli Exclusion Principle violation study, a Pb conductor was installed around the cryostat housing the HPGe crystal. Informed by \textsc{Geant4}-based~\cite{GEANT4:2002zbu, Allison:2016lfl} Monte Carlo (MC) simulations, detailed in \ref{sec:efficiency_sims}, and taking into account the mechanical feasibility of the setup, as well as safeguards against heat dissipation from the conductor to the cooled detector, a hollow cylinder geometry was selected. An inner diameter of \SI{122}{mm} ensures a \SI{1}{cm} lateral distance between the cryostat and the tripartite Pb sheet forming the cylinder. A Pb thickness of \SI{1}{mm} was chosen, a value for which the simulated detection efficiency shows a plateau for a given current density. 

\begin{figure}[htb]
	\centering
	\includegraphics[width=0.99\linewidth]{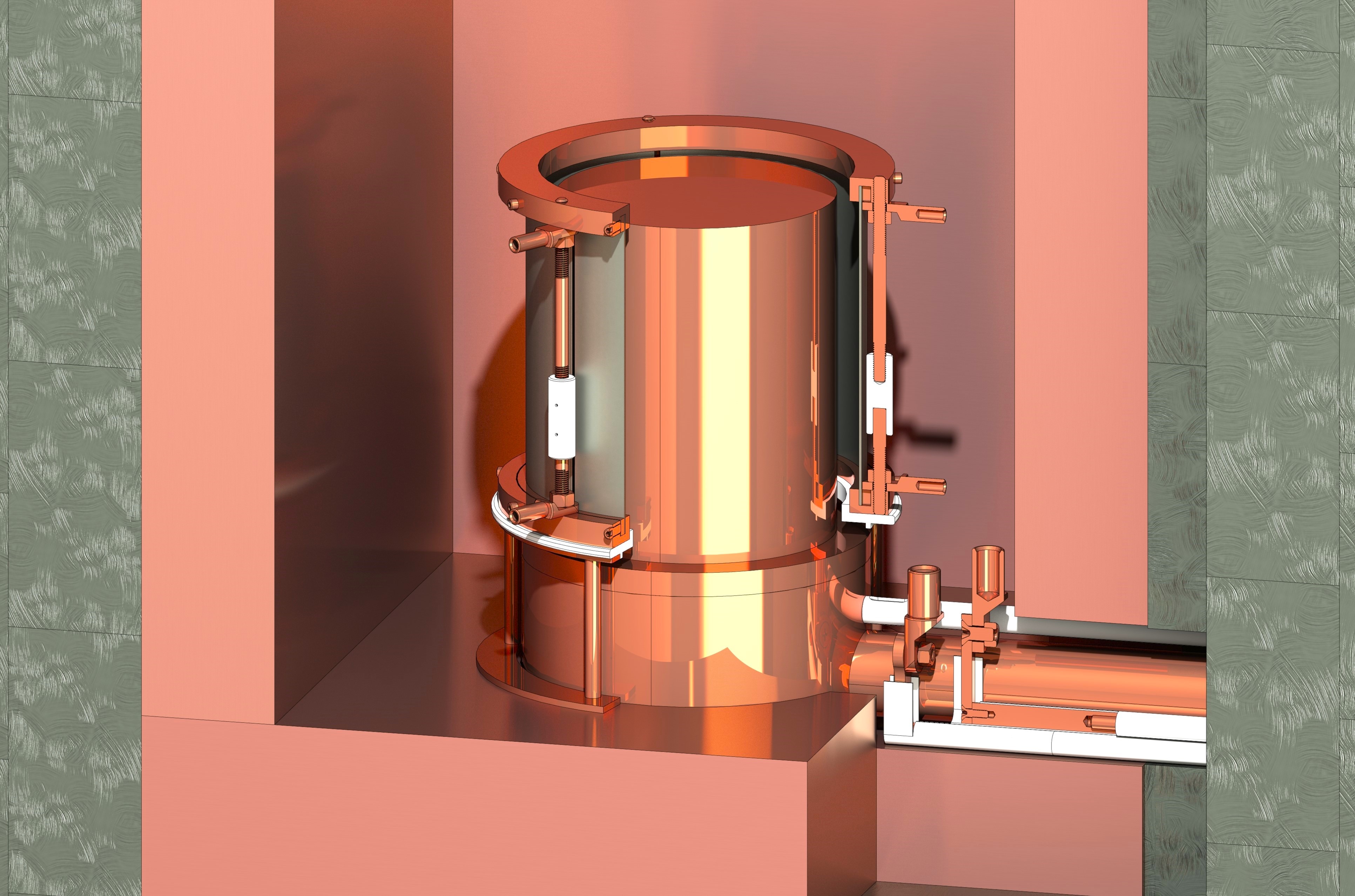}
	\caption{Schematic view of the setup for the PEP violation studies. The gray \SI{1}{mm}-thick Pb conductor cylinder is spanned by an OFHC Cu and PTFE holder structure around the cryostat of the HPGe crystal inside the sample cavity. The electric current is fed through the shield along the coldfinger with two segmented Cu rods, insulated by \SI{0.5}{mm}-thin PTFE sheets. The three \SI{10}{mm^2} cables contacting each electrode are not shown.} 
	\label{fig:setup_render}
\end{figure}

Two Cu rings with clamping elements and set screws compressing the upper and lower ends of the Pb pieces enable the electrical contact of the Pb cylinder with an effective length between the rings of \SI{10.8}{cm}. Three Cu pillars with Polytetrafluoroethylene (PTFE) insulation insets mechanically support the structure, which is additionally elevated by a light Cu pedestal with an insulating PTFE disk, such that the uncovered Pb extends vertically from the top of the Cu cryostat to about \SI{2}{cm} below the HPGe crystal. A schematic view of the setup is shown in \autoref{fig:setup_render}. 

\begin{figure}[htb]
	\centering
	\includegraphics[width=0.99\linewidth]{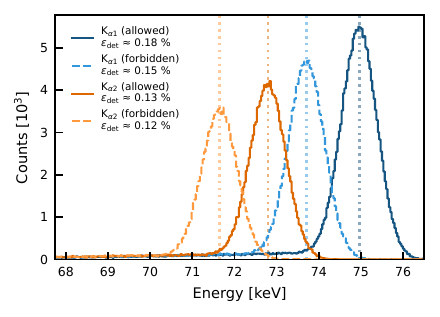}
	\caption{\textsc{Geant4} Monte Carlo simulation of $10^8$ mono-energetic X-rays each at the energies of the PEP-allowed and forbidden K$_{\alpha}$ lines in the Pb conductor (c.f., \autoref{tab:stat_model}) for the detection efficiency determination. The empirical detector resolution is applied as Gaussian smearing, and the distributions are shown in the same binning of the Gator readout MCA.} 
	\label{fig:pb_lines_sim}
\end{figure}

With this configuration, a detection efficiency of \SI{0.15}{\percent} and \SI{0.12}{\percent} for the PEP-forbidden K$_{\alpha1}$ and K$_{\alpha2}$ lines, respectively, is obtained from \textsc{Geant4} MC simulations. The simulation framework, including the implementation of the insensitive volume of the HPGe crystal, was validated with data from $^{133}$Ba and $^{228}$Th calibrations. These measurements were also used for the energy calibration of the detector, as well as for the determination of the energy-dependent resolution, which yielded values around \SI{1.0}{keV} FWHM at the energies of the investigated PEP-forbidden K$_{\alpha}$ lines. 
\autoref{fig:pb_lines_sim} illustrates the expected peaks for the PEP-allowed and forbidden K$_{\alpha1}$ and K$_{\alpha2}$ transitions, each with $10^8$ simulated primary X-rays at the respective energies. Gaussian smearing applied according to the detector resolution and binning equal to the Gator multichannel analyzer (MCA) readout demonstrates the expected peak separability. 

As discernible from \autoref{fig:pb_lines_sim}, the detection efficiency decreases rapidly towards lower energies, which is attributable to the absorption of the low-energy X-rays in the Pb itself, the Cu cryostat, and the dead layer of the HPGe crystal. Despite the comparatively low detection efficiency at energies below the typical Gator analysis region of interest, high sensitivity is still achieved through the low background of the experiment. 

To ensure low background rates from the newly installed components, in addition to the original background rates inherent to the Gator facility, low-activity materials were selected for their construction. The Pb sheets were custom cast and rolled from raw material with an activity of $< \SI{0.2}{Bq/kg}$, as reported by the manufacturer~\cite{lemerpaxLeadPreferredRadiation2024}. The Cu elements were machined from spare oxygen-free high-conductivity (OFHC) Cu from the XENONnT photomultiplier tube array support plates~\cite{XENON:2021mrg}, and the PTFE raw material stems from the XENON1T reflector plates~\cite{XENON:2017fdb}. The cables routed to the setup inside the Gator sample cavity underwent prior radioassay with Gator itself. Overall, the mass of auxiliary components besides the Pb was minimized as to lower its contribution to the radioactive background, while still ensuring mechanical stability, as well as sufficient conductor cross-sectional and surface areas to suppress heat-up from the high current. The assessment of both properties, structural rigidity and thermal safety, is detailed in \ref{sec:stability_thermal}.

\section{Measurements}
\label{sec:measurements}

The data analyzed in this work was acquired with the Gator HPGe detector between April and August 2023. A direct current of \SI{40}{A} from a current-stabilized Agilent N5761A Power Supply was passed through the Pb foil, selected based on limitations inferred from the thermal studies detailed in \ref{sec:stability_thermal}. 
One period of current-on data taking is parenthesized by two measurement phases with disabled current, serving as a background-only sample for the data analysis. The pre-amplified signals from the detector are recorded through an Ortec Model 672 spectroscopy amplifier and a self-triggering Ortec Model ASPEC-927 dual multichannel buffer with acquisition intervals of \SI{4}{h} per data set. 

Data sets acquired during the semiweekly refills of the liquid nitrogen dewar for the cold finger were found to be dominated by refill-induced noise and are thus excluded from the analysis. This data cleaning, further described in Ref.~\cite{Araujo:2022kip}, results in a live time reduction by less than \SI{5}{\percent}, yielding cumulative current-on and current-off live times of \SI{41.17}{d} and \SI{56.33}{d}, respectively.

The regular energy scale and resolution calibration analysis was conducted as detailed in Ref.~\cite{Araujo:2022kip}. 
The prominent \SI{81}{keV} $^{133}$Ba line, close to the PEP analysis ROI, exhibited maximum variations of the estimated position by about $0.1\,\sigma$. The corresponding estimated peak resolution $\sigma$ was temporally constant within $\sim \SI{1}{\percent}$. Hence, the energy scale and resolution can be considered stable and possible systematic effects on the analysis neglected.

\begin{figure}[h]
\centering
\includegraphics[width=0.99\linewidth]{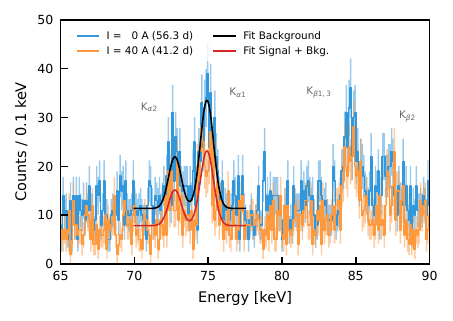} 
\caption{Data without (blue) and with (orange) current, in the \SIrange{65}{90}{keV} region acquired by the Gator HPGe detector between April and August 2023. The current was circulated on the Pb target at \SI{40}{A}. The Pb K$_{\alpha_1}$ and K$_{\alpha_2}$ lines are indicated, as well as the K$_{\beta_{1-3}}$ and K$_{\beta_2}$. 
The corresponding best-fit results for the K$_{\alpha}$ lines are overlaid.}
\label{fig:calibrated_data}
\end{figure}

The energy-calibrated spectra are shown in \autoref{fig:calibrated_data}, with and without current. Due to electromagnetic screening resulting from the additional electron on the 1s level, the forbidden transitions are expected to have a lower energy than the standard transitions. The values for the Pb K$_{\alpha_1}$ and K$_{\alpha_2}$ are given in~\autoref{tab:stat_model}.
The count rate in the energy range \SIrange[]{65}{90}{keV} with the installed setup and disabled current is, on average, increased by a factor of \SI{6.7(5)}{} compared to the rate acquired with an empty cavity. 

\section{Analysis and results}
\label{sec:results_discussion}

For this type II measurement with Gator, new fermions are introduced into the system via the \SI{40}{A} direct current. 

We have analyzed the data with a Bayesian and a frequentist approach, utilizing both the signal and background regions. In the following subsections, we detail the statistical model used for the analysis and the obtained results.

\subsection{Statistical model}

The spectral shape is described in the same way in both regions for the background: a first-degree polynomial accounts for the continuum component, and the Pb K$_{\alpha_1}$, K$_{\alpha_2}$ lines are described by Gaussian distributions. The PEP-violating (PEPV) lines, present only in the signal region, are also represented by Gaussian distributions with the same resolution as their corresponding standard transitions and their positions at the expected energies of the forbidden lines. \autoref{tab:stat_model} shows the expected positions of the K$_{\alpha_1}$ and K$_{\alpha_2}$ lines for both the standard and forbidden transitions.

\begin{table}[h]
\centering
\begin{tabular}{|c|c|c|}
\hline
\textbf{Transition} & \textbf{EM energy} & \textbf{PEPV energy} \\ \hline
1s - 2p${}_{3/2}$ K$_{\alpha1}$  & \SI{74.961}{keV} & \SI{73.713}{keV} \\ \hline
1s - 2p${}_{1/2}$ K$_{\alpha2}$  & \SI{72.798}{keV} & \SI{71.652}{keV} \\ \hline
\end{tabular}
\caption{Centers of the K$_{\alpha_1}$ and K$_{\alpha_2}$ in the electromagnetic (EM) case and the Pauli violating one~\cite{piscicchia2022strongest}.}
\label{tab:stat_model}
\end{table}

In this model, we call $\bm{\theta}=(\theta_1,\theta_2,\theta_3,\theta_4,\theta_5)$ the vector of the parameters describing the shape of the spectrum, namely, the center and resolution of the K$_{\alpha_1}$ and K$_{\alpha_2}$ lines, and the slope of the polynomial, respectively. The yields of each distribution are represented by $\bm{y}=(y_1,y_2,y_3)$, where $y_1$ and $y_2$ are the yields of the K$_{\alpha_1}$ and K$_{\alpha_2}$ lines, and $y_3$ is the yield of the continuum background. 
The number of detected forbidden K$_{\alpha_i}$, $i=1,2$, transitions is decomposed as $y_{S_i} = \mathcal{S} \cdot \epsilon_i \cdot BR_i$. Here, $\epsilon_i$ is the detection efficiency at the given energy (derived by Monte Carlo simulations, $\epsilon_1=0.0015$ and $\epsilon_2=0.0012$), and $BR_i$ is the branching ratio of the transition (0.47 for the $K_{\alpha_1}$ and 0.23 for the $K_{\alpha_2}$). The remaining free parameter $\mathcal{S}$, which is shared by construction among the two investigated forbidden lines, defines the parameter of interest. 
The model for the signal region then reads:
\begin{align}
\begin{split}
\mathcal{F}^{wc}(\bm{\theta},\bm{y},\mathcal{S}) = &\ y_1 \cdot K_{\alpha_1}(\theta_1,\theta_2)+ y_2 \cdot K_{\alpha_2}(\theta_3,\theta_4)  \\
&+ y_3 \cdot \mathrm{Pol}(\theta_5) \\
&+ \mathcal{S}\cdot \epsilon_1\cdot BR_1 \cdot \mathrm{{PEPV}_1}(\theta_2)  \\
&+ \mathcal{S}\cdot \epsilon_2\cdot BR_2  \cdot \mathrm{{PEPV_2}}(\theta_4). 
\end{split}
\end{align}

In the background region, the model reads instead:
\begin{align}
\begin{split}
\mathcal{F}^{woc}(\bm{\theta},\bm{y}) = &\ y_1 \cdot K_{\alpha_1}(\theta_1,\theta_2) + y_2 \cdot K_{\alpha_2}(\theta_3,\theta_4)\\ 
&+ y_3 \cdot \mathrm{Pol}(\theta_5).
\end{split}
\end{align}
With these ingredients, we can now proceed to write the likelihood
\begin{align}
\begin{split}
\mathcal{L}(\mathcal{D}^{wc},\mathcal{D}^{woc}|\bm{\theta},\bm{y},\mathcal{S}) = &\
\text{Poiss}(\mathcal{D}^{wc} | \mathcal{F}^{wc}(\bm{\theta},\bm{y},\mathcal{S}) ) \\
&\cdot \text{Poiss}(\mathcal{D}^{woc} | \mathcal{F}^{woc}(\bm{\theta},\bm{y} \cdot \mathcal{R})). 
\end{split}
\end{align}
Here, $\mathcal{D}$ is the data in the signal ($\mathcal{D}^{wc}$) and control ($\mathcal{D}^{woc}$) region, and Poiss($\mathcal{D}|\mathcal{F}$) is the Poisson probability of observing $\mathcal{D}$ given the expectation $\mathcal{F}$, where $\mathcal{R}$ is a normalization factor. In similar PEPV searches~\cite{napolitano2022testing}, this accounts for the different live times of the two spectra. The Gator detector, however, is also sensitive to time-dependent backgrounds originating mainly from isotopes with decay times on the same order as the measurement time. To account for  potential systematic effects, $\mathcal{R}$ is obtained by dividing the integral count rates in the two regions (estimated from a fit to the rates) as follows:

\begin{equation}
\mathcal{R} = \int_{\mathcal{D}^{wc}} \int_{E} f \cdot dt \cdot dE / \int_{\mathcal{D}^{woc}} \int_{E} f  \cdot dt \cdot dE,
\end{equation}
where $f$ is the rate fit and $E$ is the energy region \SIrange{35}{500}{keV} excluding the region of interest. We can trivially see that if $f$ is constant, $\mathcal{R}=t^{wc}/t^{woc}$, thus recovering the case of the time-independent background.
The parameters $\bm{\theta}$ and $\mathcal{R}$ are constrained via prior distributions (Bayesian approach) or penalty terms (frequentist approach). The energy scale and resolution parameters are constrained through a $^{133}$Ba calibration. Finally, the parameter of interest $\mathcal{S}$ is constrained by a flat prior.

With the statistical model defined through the likelihood, we can now proceed to the analysis results. For the Bayesian treatment, the posterior probability distribution on the parameter of interest $\mathcal{S}$ is obtained by sampling via a Markov Chain Monte Carlo (MCMC) algorithm with BAT.jl~\cite{schulz2021bat,caldwell2020integration,hafych2022parallelizing}, from which the upper limit is extracted.
 
For the (modified) frequentist treatment, we use the profile likelihood and the CL$_s$ method~\cite{read2002presentation,cowan2011asymptotic} to construct the one-sided test statistic and obtain the confidence level.

In both cases, we use a \SI{90}{\percent} confidence level for the upper limit.

\subsection{Results}

\begin{figure}[h]
\centering
\includegraphics[width=0.99\linewidth]{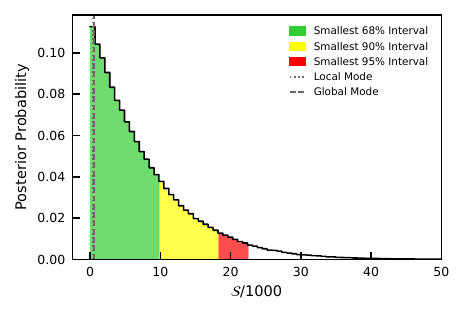}
\caption{Marginalized posterior distribution of the parameter of interest $\mathcal{S}$/1000, obtained from the Bayesian analysis. The scaling with a factor of 1000 is introduced to avoid numerical instabilities when sampling the posterior distribution. The \SI{90}{\percent} upper limit is indicated by the yellow band.}
\label{fig:posterior}
\end{figure}

In \autoref{fig:posterior}, we show the marginalized posterior distribution of the parameter of interest $\mathcal{S}$, obtained from the Bayesian analysis.  

We obtain the \SI{90}{\percent} upper limit on the parameter of interest $\mathcal{S}$ to be: 
\begin{align}
\mathcal{S} < 18000 & \textrm{ events (Bayesian)} \\
\mathcal{S} < 21000 & \textrm{ events (frequentist)} 
\end{align}

\begin{table}[h]
    \centering
    \begin{tabular}{|c|c|c|}
    \hline
    \textbf{Parameter} & \textbf{Value} & \textbf{Unit} \\ \hline
    Elementary charge ($e$) & $\SI{1.6e-19}{}$ & C \\ \hline
    Distance  ($D$) & $10.8$ & cm \\ \hline
    Scattering length  ($\mu$) & $\SI{2.34e-7}{}$ & cm \\ \hline
    Current ($I$) & 40 & A \\ \hline
    Time with current ($t$) & $41.17 \times 86400$ & s \\ \hline
    Capture probability ($P$) & 0.009 & - \\ \hline
    \end{tabular}
    \caption{Detailed values of the parameters used to derive the upper limit on $\beta^2/2$.}
    \label{tab:detailed_parameters} 
\end{table}

With the upper limit on $\mathcal{S}$, we can now proceed to derive the limit on $\beta^2/2$:
\begin{equation}
    \centering
\beta^2/2 < \frac{\mathcal{S} \cdot e }{  \frac{D}{\mu} \cdot I \cdot t \cdot P}\label{eq:limit_bayes}
\end{equation}
where $e$ is the elementary charge, $\frac{D}{\mu}$ is the current path distance over scattering length of electrons in copper, $I$ is the current, $t$ is the time with current, and $P$ is the electron capture probability by the Pb atom. The detailed values of the parameters are shown in \autoref{tab:detailed_parameters}. Using the electron diffusion model~\cite{ramberg1990experimental,shi2018experimental,milotti2018importance}, we obtain the \SI{90}{\percent} upper limit on $\beta^2/2$:

\begin{align}
\beta^2/2 < \SI{4.8e-29}{} & \textrm{ (Bayesian)} \\
\beta^2/2 < \SI{5.7e-29}{} & \textrm{ (frequentist)}.
\end{align}

\section{Conclusions and outlook}
\label{sec:conclusions}

This work presents an improvement in testing the limits of the Pauli Exclusion Principle violation probability as a function of the atomic number. Local Quantum Field Theories which allow slight deviations from Fermi and Bose statistics are subject to a stringent superselection rule \cite{superselection,amado1980comments}, first formulated by Messiah and Greenberg, which prevents transitions from PEP-standard to PEP-anomalous states. We report on the results of a dedicated experiment performed with the ultra-low background Gator facility \cite{Baudis:2011am,Araujo:2022kip}, exploiting a p-type coaxial HPGe detector operated at LNGS. The measurement searches for PEP-violating atomic transitions in Pb, fulfilling the MG superselection rule. 
The upper bound on the PEP violation probability $\beta^2/2 < 4.8 \cdot 10^{-29}$, obtained in a Bayesian analysis, improves the previous result \cite{elliott2012improved} by more than one order of magnitude. When applying the simpler frequentist comparison adopted in \ref{sec:counting_analysis}, a slightly weaker limit is obtained, which still improves the previous result by a factor of 11. 

By using Okun's words~\cite{okun1987basis}: ``It is taken for granted usually that all atoms with a given number of protons, neutrons and electrons are identical both chemically and spectroscopically. But what is the accuracy with which we know this? [\ldots] The non-Paulian atoms could be of some bizarre cosmological origin if not all of $10^{80}$ electrons in the universe were antisymmetrized.'' This work provides an upper limit for a second element of the periodic table, lead, with a comparable sensitivity with respect to the limits obtained for copper \cite{ramberg1990experimental,napolitano2022testing}, respecting the MG superselection.    
Upgrades to the experimental setup may further improve the current constraints and allow for investigating new theoretical scenarios.

\appendix
\section{Detection efficiency simulations}
\label{sec:efficiency_sims}

The design of the Pb conductor in the experimental setup was informed by \textsc{Geant4} Monte Carlo simulations of the PEP-violating Pb K$_{\alpha_1}$ X-rays. Detection efficiencies for different Pb target geometries and thicknesses were studied.

\begin{figure}[htb]
	\centering
    \includegraphics[width=0.99\linewidth]{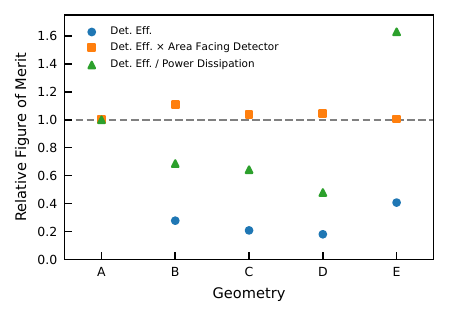}%
    \caption{Assessment of the impact of the conductor geometry at a given thickness, here taken as \SI{25}{\mu m} for facilitated simulations. Geometry A corresponds to a square plate on top of the Cu cryostat with lateral extensions equal to the end cap (i.e., the upper cryostat cylinder segment; c.f.~\autoref{fig:gator_facility}) diameter; all other geometries represent hollow cylinders. Shapes C and D have inner diameters equal to the Cu cryostat base (i.e., the lower, wider cylinder segment) diameter and span vertically from the cryostat end cap upper edge to the top of the cryostat base or the coldfinger, respectively. For realization B, the inner diameter of geometry C is shrunk to the cryostat end cap diameter, and shape E additionally shortens this modified cylinder to only extend vertically along the length of the HPGe crystal.}
    \label{fig:pep_setup_sim_geom}
\end{figure}

The geometry investigation is illustrated in \autoref{fig:pep_setup_sim_geom}. 
In terms of the absolute detection efficiency $\varepsilon$, the top plate geometry (A) yields the highest values due to the lateral crystal holder structure, and $\varepsilon$ is particularly low for larger inner Pb cylinder radii. However, the actual sensitivity depends not only on this quantity but also on the number of electron captures, which is driven by the current and its path length through the conductor. Hence, at a given conductor thickness and current density, the product of detection efficiency and conductor area facing the detector constitutes a more representative figure of merit. From this perspective, the investigated geometries exhibit performance deviations of at most $\sim \SI{10}{\percent}$, which are, hence, subdominant in the setup design considerations. 

Additional attention is further put on the power and, thus, heat dissipation in the Pb conductor, which might radiate toward the HPGe crystal. It can be seen that excessive Pb cylinder elongations may result in a much stronger impact on the heat dissipation than they gain in sensitivity, as especially visible between geometries C and D. For this reason, adequate spacer structures for the positioning of a reasonably long Pb cylinder near the HPGe crystal is required.

\begin{figure}[htb]
	\centering
	\includegraphics[width=0.99\linewidth]{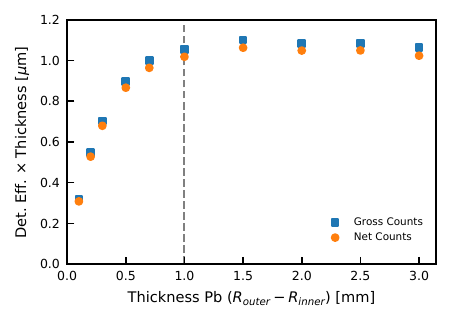}
	\caption{Product of the simulated gross (all counts at the full absorption peak energy) and net (Compton-subtracted counts) detection efficiencies with the conductor thickness as a function of the latter factor, defined as the difference between outer and inner Pb cylinder radius, for geometry D.} 
	\label{fig:pep_setup_sim_thickness}
\end{figure}

The investigation of the dependence of the detection efficiency on the conductor thickness is exemplarily depicted in \autoref{fig:pep_setup_sim_thickness} for geometry D of \autoref{fig:pep_setup_sim_geom}. The weighting of the detection efficiency with the target thickness reflects the scaling of the integral current at a given current density with the conductor cross-section. It further illustrates the X-ray gain from additional lateral material, which appears to fade due to self-absorption in the Pb around thicknesses of $\sim \SI{1}{mm}$, where the shown distributions start plateauing and which is hence used for the setup. 

\section{Stability and thermal tests setup}
\label{sec:stability_thermal}

\begin{figure}
    \centering
    \includegraphics[width=0.99\linewidth]{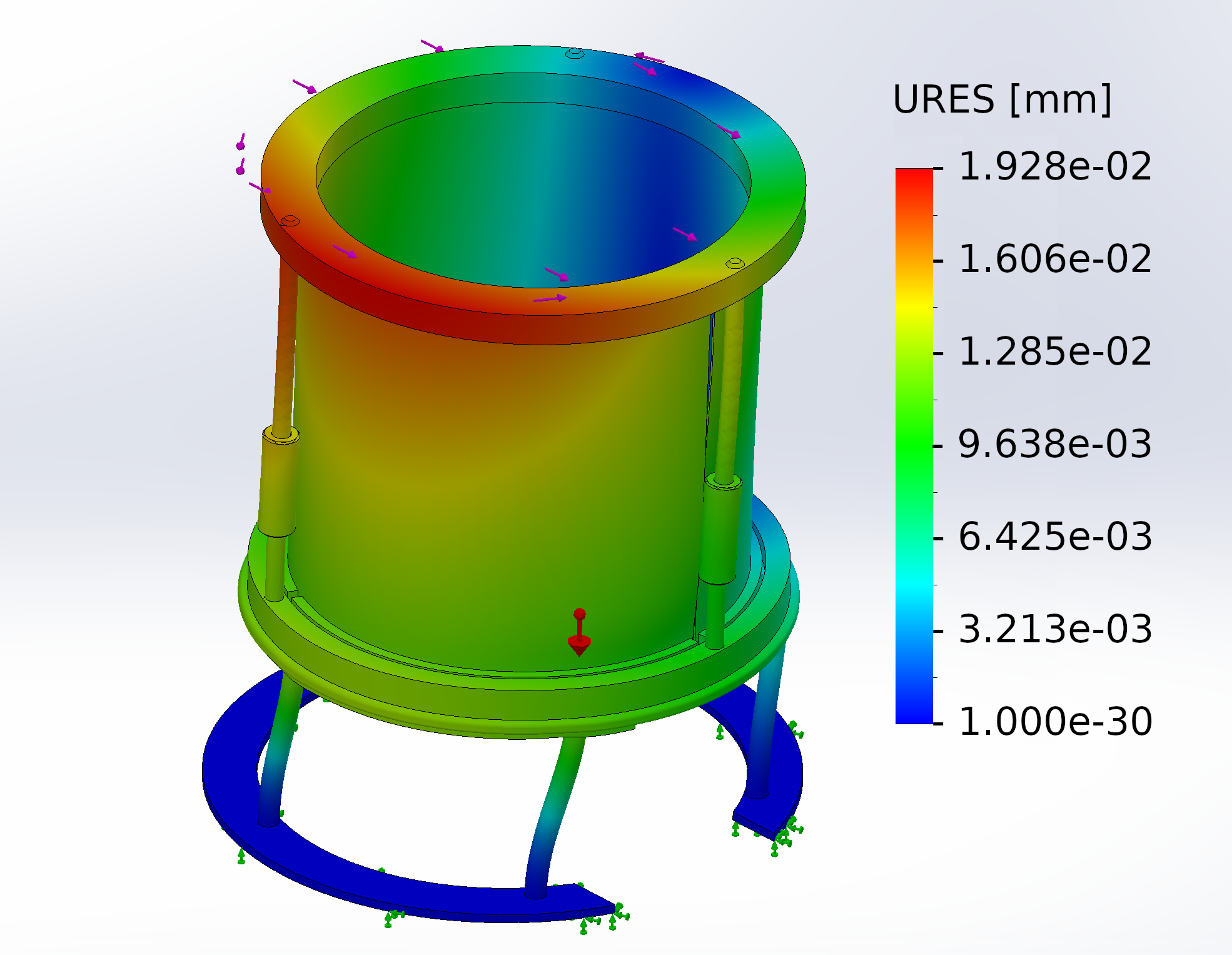}
    \caption{Mechanical assessment of the Gator PEP violation study setup: Exemplary SolidWorks stress simulation for a \SI{10}{N} horizontal force and \SI{1}{Nm} torque applied to the top Cu ring in addition to the gravitational force acting on the entire setup. The coloring encodes the resultant static displacement (URES) in mm, which can be considered negligible.}
    \label{fig:stability}
\end{figure}

The structural rigidity of the setup was verified through SolidWorks~\cite{dassaultsystemessolidworkscorporationSOLIDWORKSSimulation} simulations for different load scenarios and parameters such as stress, strain, buckling, and displacement, as exemplified in \autoref{fig:stability}. Even under the extreme assumption of a \SI{10}{N} horizontal force and a \SI{1}{Nm} torque on the top Cu ring, in addition to the gravitational force acting on the entire setup, a minimum safety factor of $\sim 40$ could be maintained.

\begin{figure}
    \centering
    \includegraphics[width=0.99\linewidth]{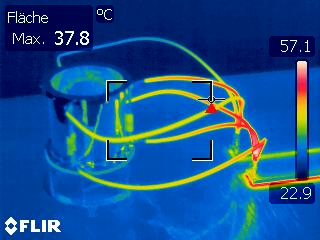}
    \caption{Thermal assessment of the Gator PEP violation study setup: Infrared image of the setup, assembled at UZH for test purposes, with a direct current of \SI{100}{A} applied.}
    \label{fig:flir}
\end{figure}

Thermal studies with currents up to \SI{100}{A} were conducted, both with an initial simplified flat geometry and the final setup assembly prior to installation in Gator. A Teledyne FLIR infrared camera image of the latter is shown in \autoref{fig:flir}. The highest local temperatures were observed on the cable junctions in the custom OFHC Cu cable lugs, likely due to the contact resistance and reduced area for heat dissipation. In order to ensure sufficient flexibility in the confined space of the Gator sample cavity, as well as to allow for a larger surface area, cabling inside the enclosure for each electrode was realized by three cables of \SI{10}{mm^2} conductor cross-section each with minimal insulation. The current routing through the shielding layers alongside the coldfinger is realized with an \SI{8}{mm}-diameter OFHC Cu rod with a five-piece segmentation, enforced by the limited installation clearance, and \SI{0.5}{mm}-thin PTFE sheet wrapping for electrical insulation. The feedthrough in the Gator enclosure top plate is realized via a high-current D-subminiature connector with \SI{40}{A}-rating per pin. During the commissioning of the setup in Gator, heat-up studies over several hours in the confined environment of the facility with regular temperature measurements at multiple control points through a contact thermometer suggested the reduction of the original target current of \SI{100}{A} to \SI{40}{A} due to safety concerns. This decision was especially driven by the fact that the LN$_2$ levelmeter had to be disabled during data taking to mitigate noise leakage into the search ROI.

\section{Counting method analysis}
\label{sec:counting_analysis}

In order to obtain an analysis method-independent estimate on the sensitivity improvement from the measurements presented in this manuscript with respect to the one in Ref.~\cite{elliott2012improved}, we show results following the same conservative analysis as used in Ref.~\cite{elliott2012improved} in this section. This method uses an approach based on the gross counts at the PEP-violating Pb X-ray lines. The integration window for both investigated lines is chosen such that the ratio $\sqrt{B}/\varepsilon_{ROI}$ is minimized. At this, $B$ denotes the background from the current-off data within the window, which might be disproportionately inflated if neighboring peaks are substantially contained. The efficiency factor $\varepsilon_{ROI}$, on the other hand, expresses the peak shape fraction contained within the chosen region of interest. A scan of independent left and right window boundaries yields marginally asymmetric ROIs of \SIrange[]{73.22}{74.31}{keV} and \SIrange[]{71.11}{72.16}{keV} with $\varepsilon_{ROI,1} \approx 0.7995$ and $\varepsilon_{ROI,2} \approx 0.7895$, respectively. The product of these respective efficiency values with the branching ratios $\varepsilon_{BR,1} = 0.47$ and $\varepsilon_{BR,2} = 0.23$~\cite{elliott2012improved},
as well as the simulated detection efficiencies $\varepsilon_{det,i}$, yields the total efficiencies, $\varepsilon_{tot,i}$, for the investigated PEP-forbidden lines.

After live time correction of the current-off counts for both PEP-forbidden lines, efficiency-uncorrected count differences between current-on and off of $N_1 = \SI{-16(14)}{}$ and $N_2 = \SI{-19(12)}{}$
are obtained. This corresponds to current-on rates lower by $\SI{9(8)}{\percent}$ and $\SI{14(9)}{\percent}$, respectively, than in the data with disabled current, which is consistent within the large uncertainties
with the reduction expected from the background time dependence. Nominally, however, it constitutes an additional local underfluctuation at the investigated ROIs. This underfluctuation is not significant compared to the global fluctuations, which is why no background time dependence correction is applied to avoid biasing the limit setting. Correcting for the deviating current-off live time
and dividing the measured counts by the total efficiencies $\varepsilon_{tot,i}$, a weighted average net count of
$\SI{-3.16(1.65)e4}{}$ for a current-on live time is obtained, where for the recorded counts, a Poisson error is assumed. In order to constrain these downward fluctuations and background time dependence, and thus mitigate spurious exclusion, a conservative zero net count is assumed for the limit setting. Therefore, the further considerations are based on a $3\sigma$ upper limit of $N_{3\sigma}/\varepsilon_{tot} \approx 3 \times \SI{1.65e4}{}$.

The $3\sigma$ upper limit on the PEP violation probability can then be calculated as
\begin{equation}
    \centering
\beta^2/2 < \frac{N_{3\sigma} \cdot e }{\varepsilon_{tot} \cdot \frac{D}{\mu} \cdot I \cdot t \cdot P},
\end{equation}
with variable names and values as in \autoref{eq:limit_bayes}. This yields $\beta^2/2 < \SI{1.3e-28}{}$, which constitutes a factor $\sim 11$ improvement compared to the previous best limit for lead with
the Ramberg-Snow technique in Ref.~\cite{elliott2012improved}, employing the same analysis method.

\begin{acknowledgements}

This work was supported by the European Research Council (ERC) under the European Union’s Horizon 2020 research and innovation programme, grant agreement No. 742789, and by the SNF grant 20FL20-201437. We thank the electronics and mechanical workshops in the UZH Physics Department for their continuous support.

We acknowledge support from the Foundational Questions Institute and Fetzer Franklin Fund, a donor advised fund of Silicon Valley Community Foundation (Grants No. FQXi-RFP-CPW-2008 and FQXi-MGA-2102), and from the H2020 FET TEQ (Grant No. 766900). This publication was made possible through the support of Grant 62099 from the John Templeton Foundation. The opinions expressed in this publication are those of the authors and do not necessarily reflect the views of the John Templeton Foundation.

K.P. acknowledges support from the Centro Ricerche Enrico Fermi - Museo Storico della Fisica e Centro Studi e Ricerche ``Enrico Fermi'' (Open Problems in Quantum Mechanics project).

We thank INFN for supporting the research presented in this article and, in particular, LNGS, its Director, Ezio Previtali, and the LNGS
staff.

\end{acknowledgements}

\bibliographystyle{JHEP}
\bibliography{gator_vip}  

\end{document}